\documentclass[prl,twocolumn,superscriptaddress]{revtex4-1}

\usepackage[letterpaper,top=1in, bottom=1.22in, left=1.40in, right=0.850in]{geometry}
\usepackage{setspace}
\usepackage{amsfonts,amssymb,amsmath}            % for math symbols.
\usepackage{amsbsy,mathrsfs}     %for no extensions (chatper 4)
\usepackage[normalem]{ulem}
\usepackage{bm}
\usepackage{subfigure}
\usepackage{booktabs}
\usepackage{epsfig,graphics}
\usepackage{amsthm}
\usepackage{textcomp}
\usepackage{multirow} 
\usepackage{dsfont}
\usepackage{bbold}
\usepackage[utf8x]{inputenc}

\newcommand{\ket}[1]{\vert#1\rangle}
\newcommand{\bra}[1]{\langle#1\vert}

\def\opone{\leavevmode\hbox{\small1\kern-3.8pt\normalsize1}}

\begin{document}

%\widowpenalty=20000
%\clubpenalty=20000

\title{A Flexible Source of Non-Degenerate Entangled Photons Based on a Two-Crystal Sagnac Interferometer}
\author{Terence E. Stuart}
\affiliation{Institute for Quantum Information Science and Department of Physics \& Astronomy, University of Calgary,  Calgary, Alberta T2N 1N4, Canada}
\author{Joshua A. Slater}
\affiliation{Institute for Quantum Information Science and Department of Physics \& Astronomy, University of Calgary,  Calgary, Alberta T2N 1N4, Canada}
\author{F\'{e}lix Bussi\`{e}res}
\affiliation{Institute for Quantum Information Science and Department of Physics \& Astronomy, University of Calgary,  Calgary, Alberta T2N 1N4, Canada}
\affiliation{GAP-Optique, Universit\'{e} de Gen\`{e}ve, Geneva, Switzerland, CH-1211 Genève 4}
\author{Wolfgang Tittel}
\affiliation{Institute for Quantum Information Science and Department of Physics \& Astronomy, University of Calgary,  Calgary, Alberta T2N 1N4, Canada}

\begin{abstract}
Sources of entangled photon pairs are a key component in both fundamental tests of quantum theory and practical applications such as quantum key distribution and quantum computing.  In this work, we describe and characterize a source of polarization entangled photon pairs based on two spontaneous parametric down-conversion (SPDC) crystals in a Sagnac interferometer.  Our source is compact and produces high-quality entangled states in a very flexible manner.  The wavelengths of the photon pairs, around 810 and 1550 nm, the phase between orthogonal components of the entangled state, and the tangle of the state are all independently adjustable. In addition to presenting basic characterization data, we present experimental violations of CHSH and Leggett inequalities, as well as an instance of the ``beautiful'' Bell inequality, which has not previously been tested experimentally.
\end{abstract}

\date{\today}
\maketitle
\pagenumbering{arabic}     % resets page counter to one
\setcounter{page}{1}
%<<Make sure we mention advantages of our design somewhere in background>>
%<<Mention that controlled tangle is a selling point of this source.  Why would it be useful?>>
%<<Werner states: we're not really producing them!>>
%Should we include a plot of all measured spectra, either in main body or supplmenets?
%<<Spectra>>

%
%1. Bring the overlap stuff earlier in the paper.  
%2. Look at formatting of appendix
%4. Acknowledgements (after conclusion, before appendix)

%Entanglment is important -> Several high performance sources have been based on Sagnac inter... -> A problem with these is chromatic dispersion from highly separated wavelenghts -> We propose and test a novel design that addresses this problem and, in addition, is highly flexible...

\section{Introduction}
Over the last century quantum theory has fundamentally changed our understanding of the universe and continues to offer new insights into nature. Schr\"odinger described entanglement as ``\textit{the} characteristic trait of quantum mechanics''\cite{schrodinger}.  As such, it is not surprising that sources of entangled particles are a key resource in experiments that probe aspects of quantum theory~\cite{entanglement_review}.  They are also fundamental building blocks for practical applications of quantum information theory, such as quantum key distribution~\cite{Ekert91} and linear optical quantum computing~\cite{LOQCreview}.  Sources of entangled photon pairs based on SPDC in non-linear crystals~\cite{Burnham1970} are now widely used, and several high performance entanglement sources have been based on a non-linear crystal in a Sagnac interferometer thanks to this type of interferometer's intrinsic phase stability~\cite{Kim2006}. 
However, due to problems arising from chromatic dispersion in polarization optics, such sources are challenging to build if the members of the entangled pairs are generated at widely different wavelengths. One way to overcome this problem is to use periscopes instead~\cite{Periscope}. Here we resort to another approach, which is based on a Sagnac interferometer that includes two SPDC crystals. In addition to being compact and highly flexible in terms of the states it can produce, an interesting added feature is that the quality of entanglement (the tangle) can be varied in a controlled manner.  Our source has proved suitable for fundamental tests of quantum theory, some of which have not been performed before, and would also be well suited to  applications requiring transmission of entangled photons through both optical fiber and free space, e.g. for hybrid quantum networks.  

\section{Source Design}
\begin{figure*}[htp]
\centering
\includegraphics[width=0.75\textwidth]{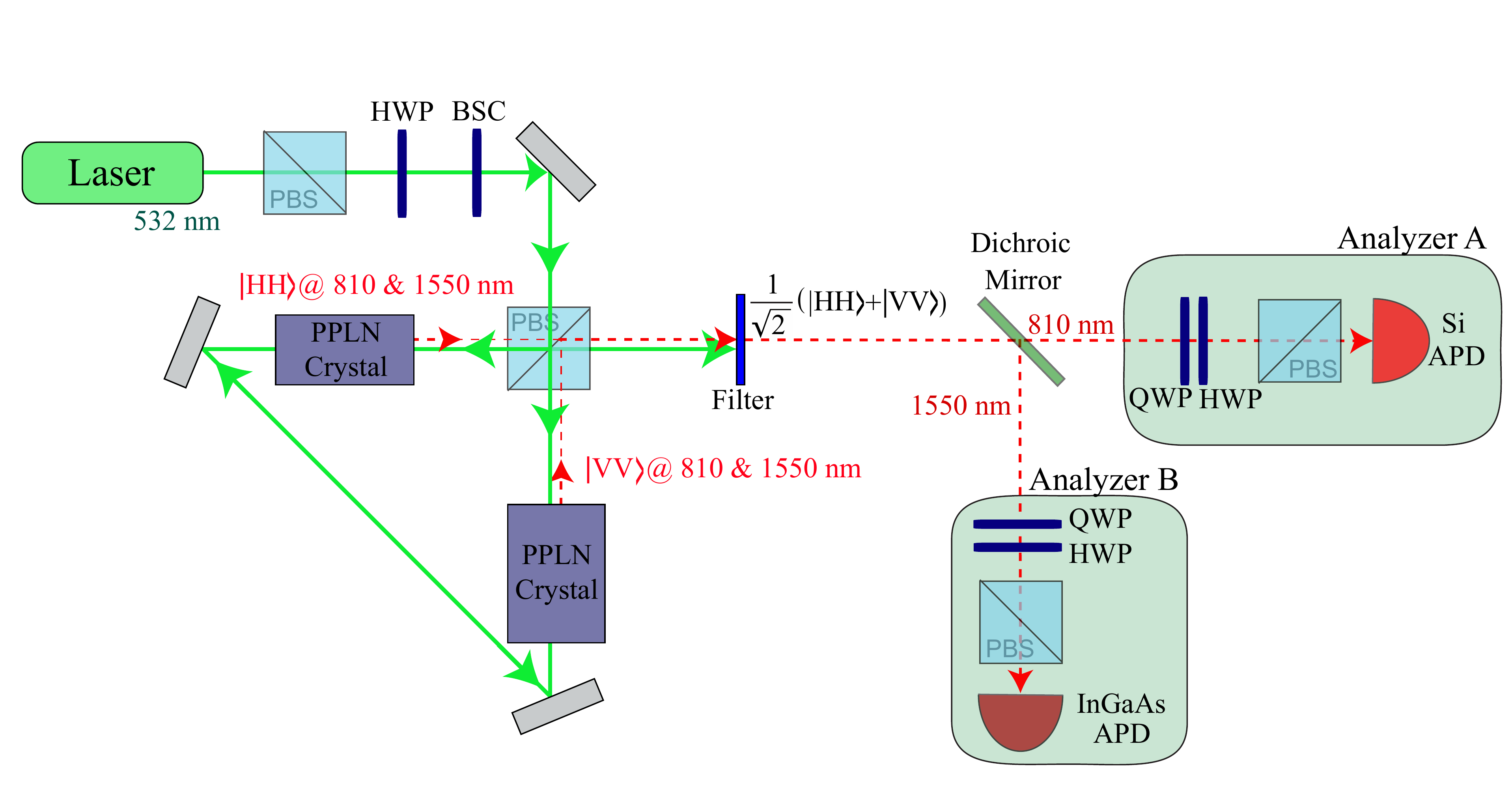}
\caption[Polarization entanglement source with qubit analyzers]{\label{fig:source}{\bf Polarization entanglement source with qubit analyzers.} Entangled states produced by the source are split according to wavelength on a dichroic mirror and distributed to analyzers A and B, which are each composed of a $\frac{\lambda}{4}$ waveplate (QWP), a $\frac{\lambda}{2}$ waveplate (HWP), polarizing beam splitter (PBS) and wavelength specific single photon detectors (Si APD and InGaAs APD). See text for details.}
\end{figure*}

Figure \ref{fig:source} shows the design of our entanglement source.  Depending on the experiment, light from a 532~nm pulsed or continuous wave laser is linearly polarized before being rotated to an equal superposition of horizontal and vertical polarizations using a $\frac{\lambda}{2}$ waveplate.  Pump light is then split into two paths by a polarizing beam splitter (PBS).  In the clockwise (CW) branch of the interferometer, horizontally polarized pump light first encounters a periodically poled lithium niobate (PPLN) crystal that is oriented to satisfy the phase matching conditions for SPDC with vertically polarized pump light.  The pump light will thus pass through this crystal without interaction because the phase matching conditions are not met at this polarization.  The second PPLN crystal encountered by pump light in this path is oriented to down-convert horizontally polarized pump light, so pairs of horizontally polarized photons at non-degenerate wavelengths of 810~nm and 1550~nm are now produced. These pairs are transmitted through the PBS and exit the source.  The counter-clockwise (CCW) path is similar, except that vertically polarized pairs are produced in the second crystal encountered and then reflected into the same output mode as the horizontal pairs from the CW path. The pump intensity is adjusted so that single photon-pair events dominate detection statistics, as evidenced by the results shown below.  Since pump light travels through both arms of the interferometer in a coherent superposition, recombining both arms on the PBS produces the entangled state,

\begin{equation}\label{eq:state}
 \ket{\Phi^\phi}=\frac{1}{\sqrt{2}}\left(\ket{HH}+e^{i \phi}\ket{VV}\right) . 
\end{equation}
 
\noindent The phase, $\phi$, is controlled using a Babinet-Soleil phase compensator (BSC) placed in front of the interferometer, which allows changing the phase between the horizontally and vertically polarized components of the pump laser. For the data collected for this article, $\phi$ was chosen to be close to zero so that the resulting state had a high fidelity with a $\ket{\Phi^+}$ Bell state.   

After the pump light is filtered out, photon pairs are separated according to wavelength by a dichroic mirror and sent to wavelength specific qubit analyzers consisting of a $\frac{\lambda}{4}$ waveplate, a $\frac{\lambda}{2}$ waveplate, a PBS, and wavelength specific detectors, as shown in Figure \ref{fig:source}.  These analyzers allow arbitrary projection measurements to be made on each of the photons.  A free running silicon avalanche photo-diode (Si APD) is used in the 810~nm photon analyzer, A.  Its output is used to trigger an Indium Galium Arsenide (InGaAs) APD used in the 1550~nm analyzer, B. Detection signals are collected using a Time-to-Digital Converter (TDC) so that coincidences between detection events can be recorded.  Using approximately $2$~mW of pump power, signal photon detections occur at a rate of approximately $20$~KHz and coincidences at a rate of approximately $500$~Hz.  The dark count rate for the Si APD is approximately $40$~Hz, and the InGaAs APD has a dark count rate of $5$x$10^{-5}$/ns.  

\section{Visibility and Quantum~State~Tomography}
%------------------------------------ Visibility, Tangle, Density Matrix ------------------------------------------------------------

Two-photon interference visibilities were assessed by performing two sets of measurements using the continuous wave pump laser. In the first measurement analyzer A (810 nm) projected onto $\ket{H}$ while the analyzer B (1550 nm) projected onto states represented on the great circle around the Bloch sphere that includes $\ket{H}$, $\ket{V}$, $\ket{+}$, and $\ket{-}$. In the second measurement, the analyzer A projects onto $\ket{+}$ and the analyzer B projects onto states represented on the great circle including $\ket{R}$, $\ket{L}$, $\ket{+}$, and $\ket{-}$. Here, $\ket{+}$ and $\ket{-}$ denote $\pm$ 45$^o$ linear polarization, and  $\ket{R}$ and $\ket{L}$ denote right and left circular polarization, respectively. Fitting the measured coincidence rates to sinusoidal functions with visibilities $V_1$ and $V_2$, we find $V_1=(99.1 \pm 0.7)\%$ and $V_2=(97.4 \pm 0.9)\%$, both being close to the maximum value of 100\%.  

Table \ref{tab:density_matrix} shows data of a typical density matrix resulting from maximum likelihood quantum state tomography (QST)~\cite{Kwiat_tomography} with a tangle~\cite{Tangle} of $\mathcal{T}=0.905$. 

\begin{table}[htp]
\centering
\subtable[\ $Re\{\rho\}$]{
\footnotesize
\centering
\begin{tabular}{r | c  c  c  c  } 
\multicolumn{5}{c}{} \\
 \multicolumn{1}{r}{} &  \multicolumn{1}{c}{$\bra{HH}$} & \multicolumn{1}{c}{$\bra{HV}$} & \multicolumn{1}{c}{$\bra{VH}$} & \multicolumn{1}{c}{$\bra{VV}$} \\ \cline{2-5}
 $\ket{HH}$ & 0.5085 & 0.0085 & -0.0151 & 0.4773  \\
 $\ket{HV}$ & 0.0085 & 0.0028 & -0.0006 & 0.0145  \\
 $\ket{VH}$ & -0.0151 & -0.0006 & 0.0038 & -0.0075  \\
 $\ket{VV}$ & 0.4773 & 0.0145 & -0.0075 & 0.4848  \\
\end{tabular}
}
\subtable[\ $Im\{\rho\}$]{
\footnotesize
\centering
\begin{tabular}{r | c  c  c  c  } 
\multicolumn{5}{c}{} \\
 \multicolumn{1}{r}{} &  \multicolumn{1}{c}{$\bra{HH}$} & \multicolumn{1}{c}{$\bra{HV}$} & \multicolumn{1}{c}{$\bra{VH}$} & \multicolumn{1}{c}{$\bra{VV}$} \\ \cline{2-5}
 $\ket{HH}$ & 0.0000 & 0.0028 & -0.0027 & -0.0337  \\
 $\ket{HV}$ & -0.0028 & 0.0000 & 0.0028 & 0.0036  \\
 $\ket{VH}$ & 0.0027 & -0.0028 & 0.0000 & -0.0045  \\
 $\ket{VV}$ & 0.0337 & -0.0036 & 0.0045 & 0.0000  \\
\end{tabular}
}
\caption[Typical Density Matrix]{\label{tab:density_matrix}{\bf Typical Density Matrix.}  Real and imaginary
  parts of the density matrix generated by maximum likelihood QST performed when the spectral overlap between SPDC crystals was optimized. The tangle is($\mathcal{T}=0.905$).}
\end{table}

%------------------------------------------------ Controlling Tangle -------------------------------------------------------- 
\section{Controlling Tangle}
In order for the entangled state produced by this source to be of high quality (i.e. to have a tangle close to 1), the spectra produced by the two SPDC crystals must match as closely as possible. Imperfectly overlapping spectra yield information that reveals in which crystal a given pair of photons was created, thus reducing the tangle of the state.  The crystals used were made by the same manufacturer, but at different times and therefore have slightly different poling periods if they are at the same temperature. By maintaining the SPDC crystals at slightly different temperatures we can select the phase-matching conditions such that the spectra of the $\ket{HH}$ and $\ket{VV}$ photon pairs are nearly indistinguishable. This changes the phase $\phi$ of the state in eq.~\ref{eq:state}, which we compensate for using the BSC. It is also possible to deliberately mismatch the spectra in a controlled way, allowing this source to produce states with an arbitrary degree of entanglement. This is done by adjusting the temperature of one PPLN crystal relative to the other, thus altering the spectrum of photons it produces and reducing the spectral overlap between pairs produced by the two SPDC crystals.  
 
\begin{figure}[htp]
\centering
\includegraphics[width=0.95\columnwidth]{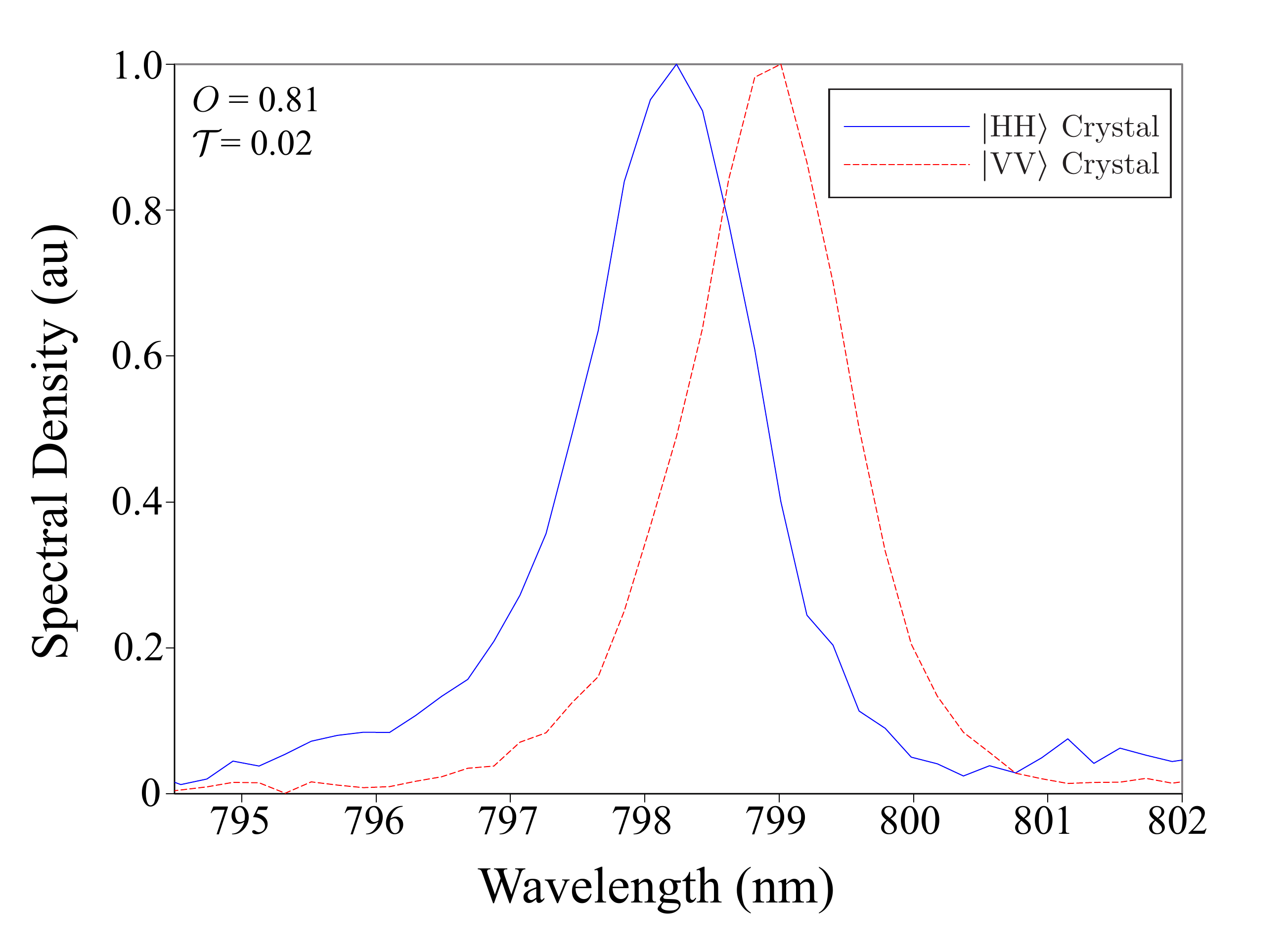}
\caption[Single photon spectra for two crystals at different temperatures]{\label{fig:spec_dist_signal_spectra}{\bf Single photon spectra for two crystals at different temperatures.} This plot shows single photon spectra gathered for $\sim 810~nm$ signal photons from the entanglement source's $\ket{VV}$ PPLN crystal at $T=165.70~^\circ C$ and from the $\ket{HH}$ PPLN crystal at $T=165.20~^\circ C$.}
\end{figure}

Figure \ref{fig:spec_dist_signal_spectra} shows two signal spectra one gathered from the $\ket{HH}$ PPLN crystal at $T=165.2~^\circ C$ and the other gathered from the $\ket{VV}$ PPLN crystal at $T=165.70~^\circ C$. For these temperatures the two spectra have incomplete overlap $O$ (see equation~\ref{eq:overlap}), and the tangle $\mathcal{T}$ of the photon pairs produced is small, but non-zero. Note that the data presented in this section has been taken with the pulsed pump; all other data has been taken with the continuous wave laser.
 
To see how tangle is related to spectral overlap, we then varied the temperature of the PPLN crystal that down-converts pump light in the CW path of our entanglement source while the other SPDC crystal's temperature was held constant.  This shifted the spectrum of the $\ket{HH}$ component of the state relative to the $\ket{VV}$ component, resulting in different degrees of spectral overlap, $O$, which we calculate as:
\begin{equation}\label{eq:overlap}
	O = \int \sqrt{S_{HH}(\lambda)} \sqrt{S_{VV}(\lambda)} d\lambda  .
\end{equation} 

\noindent where $S_{HH}(\lambda)$ is the the signal spectral density as a function of wavelength, $\lambda$, for the SPDC crystal producing $\ket{HH}$ photons pairs and $S_{VV}(\lambda)$ is the signal spectral density of the SPDC crystal producing $\ket{VV}$ photon pairs.

\begin{figure}[htp]
\centering
\includegraphics[width=0.99\columnwidth]{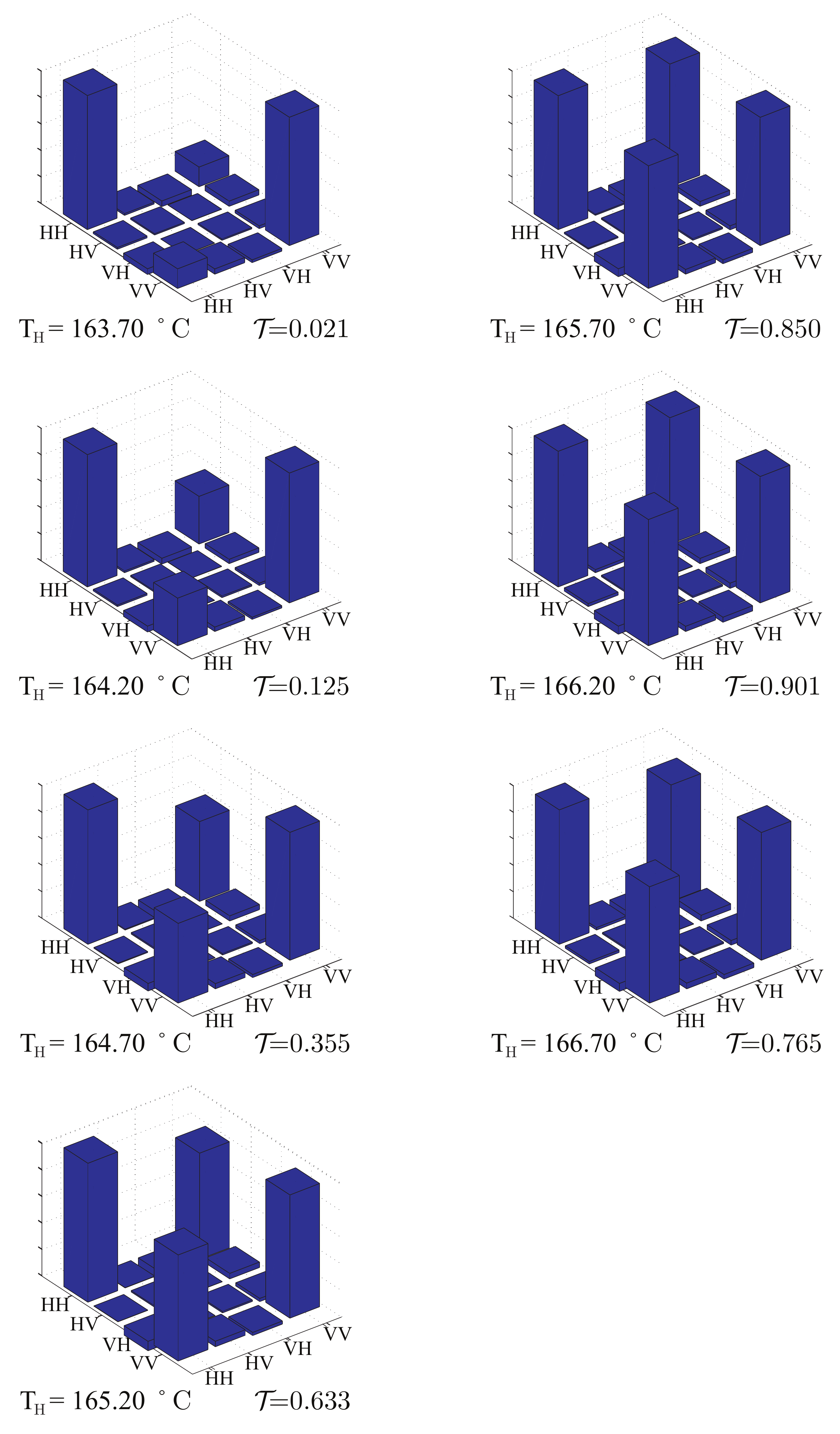}
\caption[Density matrices for different temperatures]{\label{fig:tangle_vs_temp_graphical}{\bf Density matrices for different temperatures.} This plot depicts the real components of the density matrices shown from each data point in Figure~\ref{fig:spec_dist_tangles}, ordered column wise by crystal temperature. Full density matrices for each point are detailed in Table~\ref{tab:tangle_vs_temp}. }
\end{figure}

\begin{figure}[htp]
\centering
\includegraphics[width=0.95\columnwidth]{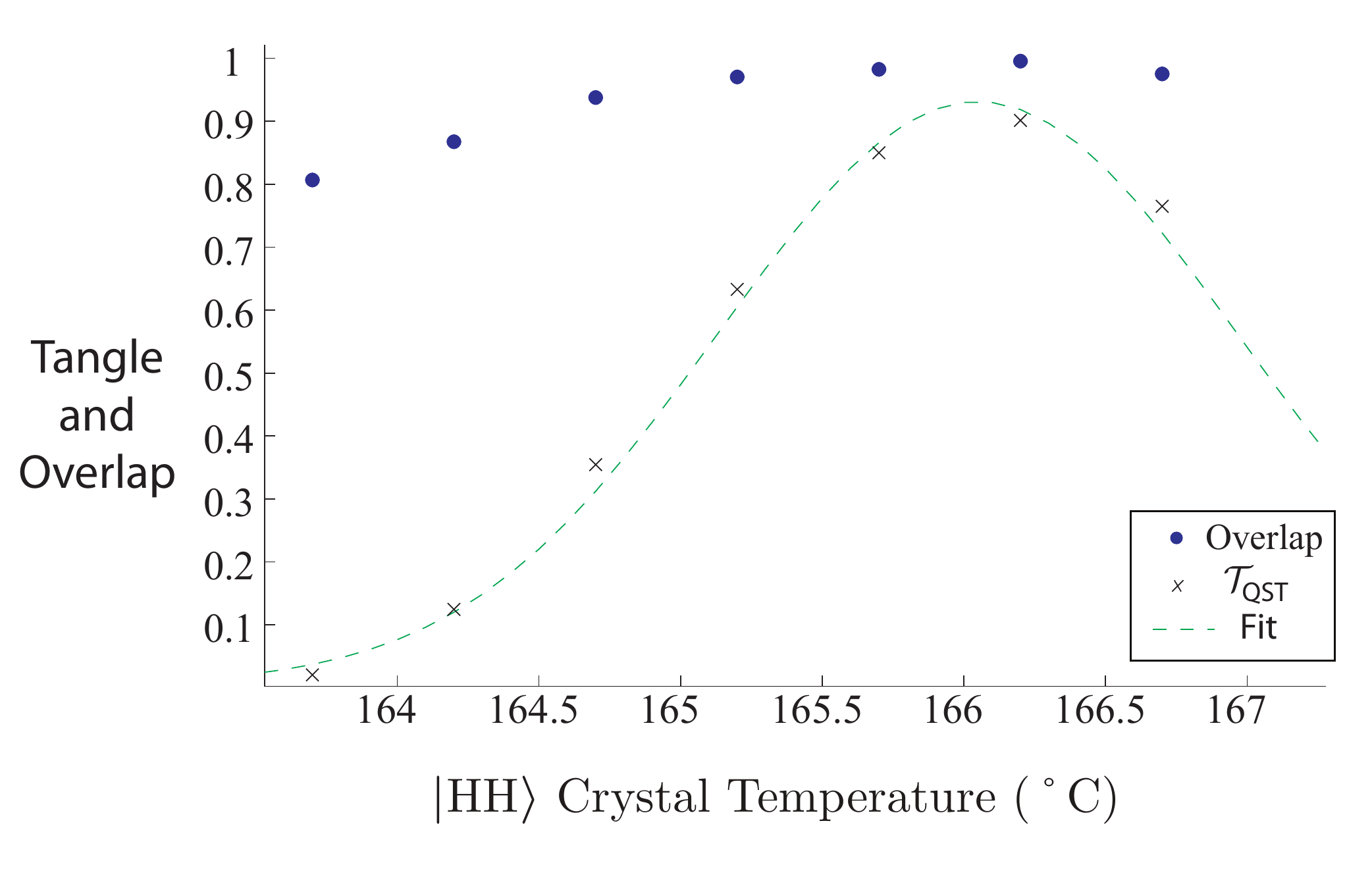}
\caption[Tangle vs spectral overlap]{\label{fig:spec_dist_tangles}{\bf Tangle vs spectral overlap.} This plot shows tangles derived from density matrices (shown in Table \ref{tab:tangle_vs_temp}) measured via QST, $\mathcal{T}_{QST}$, as the spectral overlap was changed by varying the temperature of the $\ket{HH}$ PPLN crystal.  The $\ket{VV}$ crystal's temperature was kept constant.  Also shown is the overlap, $O$, of the measured spectra.}
\end{figure}

We measured the spectrum of the signal photons from the $\ket{VV}$ SPDC crystal, which was kept at a constant temperature of $T=165.70~^\circ C$ using a temperature controlled oven that is stable to $\pm~0.01~^\circ C$.  We also measured spectra of signal photons from the $\ket{HH}$ SPDC crystal at several different temperatures. At each of these temperatures we also performed QST on the resulting bipartite states to find density matrices and associated tangles for each temperature as shown in Figure~\ref{fig:tangle_vs_temp_graphical}.  Tangle and overlap vs crystal temperature are shown in Figure \ref{fig:spec_dist_tangles}.  

%----------------------------------------------------- CHSH -------------------------------------------------------------------------

\section{Tests of CHSH Bell, ``Beautiful'' Bell, and Leggett Inequalities} 

\subsection{Bell inequalities}

\begin{figure}[htp]
\centering
\includegraphics[width=0.55\columnwidth]{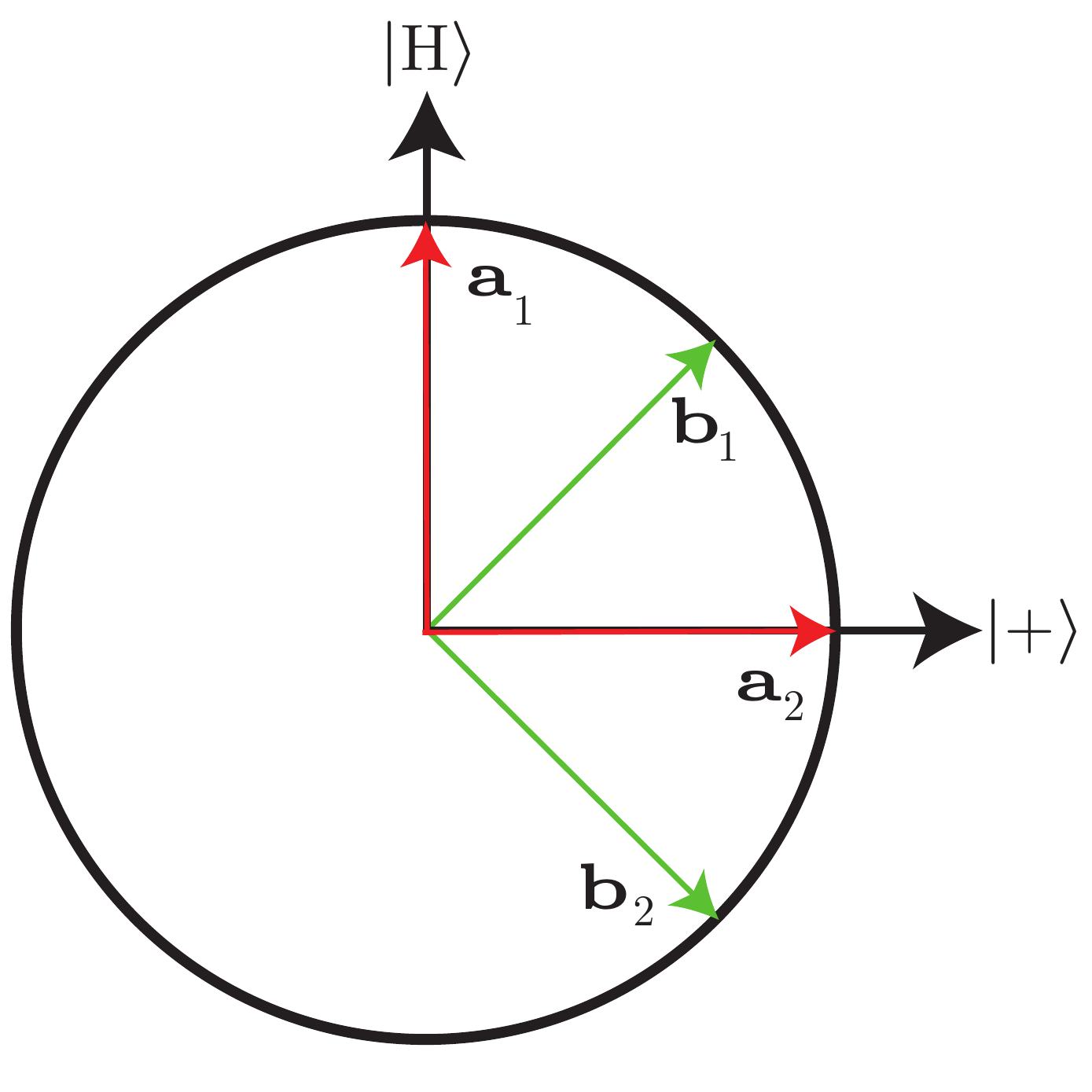}
\caption[CHSH Measurement Bases]{\label{fig:CHSH}{\bf CHSH Measurement Bases.} An optimal set of measurement bases for testing the CHSH Bell inequality when using a $\ket{\Phi^+}$ state is shown here on the equator of the Bloch sphere. Only one vector for each basis is shown.  The orthogonal vector associated with each basis is rotated by $\pi$ from the vector shown. }
\end{figure}

To assess the non-classical properties of the states produced by our source we first tested the CHSH Bell inequality~\cite{CHSH}.  A violation of this inequality demonstrates that local hidden variable (LHV) models are not adequate to describe the behaviour of the states the source is producing and demonstrates the presence of entanglement. In the CHSH inequality, Alice and Bob each measure in one of two bases, chosen uniformly and at random.  For each combination of bases, $\hat{a}_i=\{a_i,a^\perp_i\}$ and $\hat{b}_j=\{b_j,b^\perp_j\}$, Alice and Bob measure the correlation coefficient, 

\begin{align} \label{eq:CHSH_E}
E(\hat{a}_i,\hat{b}_j) = & P(a_i,b_j)+P(a^\perp_i,b^\perp_j) \\ \nonumber
 						& -P(a^\perp_i,b_j)-P(a_i,b^\perp_j) ,
\end{align}

\noindent where:

\begin{align*}
&P(a_i,b_j)=\\
&\quad\frac{C(a_i,b_j)}{C(a_i,b_j)+C(a_i^\perp,b_j)+C(a_i,b_j^\perp)+C(a_i^\perp,b_j^\perp)}
\end{align*}

\noindent and $C(a_i,b_j)$ is the number of ``coincidence'' detections observed when Alice and Bob projectively measure along basis vectors $a_i$ and $b_j$ respectively. One optimal set of bases for testing a the CHSH Bell inequality with a $\ket{\Phi^+}$ state is shown in Figure~\ref{fig:CHSH}. We then calculate the Bell S parameter as:

\begin{align}
	S =& E(\hat{a}_1,\hat{b}_1)-E(\hat{a}_1,\hat{b}_2)\nonumber\\
	&+E(\hat{a}_2,\hat{b}_1)+E(\hat{a}_2,\hat{b}_2).
\end{align}

LHV models predict that S must fall within the range: $-2\leq S \leq 2$.  Measurements made with our source (again using the continuous wave laser) produced a value of $S=2.757\pm0.008$.  The uncertainty is based on Poissonian statistics.  We note that QST yielded a density matrix with a tangle of $\mathcal{T}=0.884$ immediately before this measurement.  Based on this we would expect a maximum S parameter value of $S_{max}=2\sqrt{1+\mathcal{T}}=2.75$, which is consistent with the measured value. 

%--------------------------------------------------- Beautiful Bell -----------------------------------------------------------------

In the CHSH Bell inequality two particles, each with a Hilbert space of dimension $m=2$, are distributed to Alice and Bob. Alice makes projective measurements onto 4 states in $n=2$ bases. For an optimal violation of the bound given by the inequality, Alice chooses bases that are mutually unbiased and Bob makes {projective measurements} onto all $m^n=4$ possible intermediate states (see~\cite{belle_bell_1} {for a precise definition). An interesting question is if (and how) Bell inequalities can be constructed that a) make use of higher-dimension states or larger number of measurements made by Alice, and b) require similarly symmetric projection measurements for maximum violation. The ``beautiful'' Bell family of inequalities~\cite{belle_bell_name} was proposed by H. Bechmann-Pasquinucci and N. Gisin in 2003~\cite{belle_bell_1} and expanded upon by Gisin in 2008~\cite{belle_bell_2} in response to these questions. The authors proposed a general form of Bell inequalities, parametrized by $m$ and $n$, for which the CHSH Bell inequality is the specific case in which $m=2$ and $n=2$. The next simplest (and only) inequality in the ``beautiful'' Bell family that we can evaluate with a source of entangled qubits is the $m=3$, $n=2$ case. This inequality differs from the CHSH Bell inequality in that Alice measures in 3 bases, each spanned by two orthogonal states. Some reflection yields $m^n=2^3=8$ intermediate states that Bob needs to projectively measures onto~\cite{belle_bell_1}. The optimal measurement bases for the $m=3$, $n=2$ case are shown in Figure~\ref{fig:belle_bell_settings} -- note their highly symmetric distribution around the Bloch sphere.

The (2,3) ``beautiful'' Bell inequality  reads:

\begin{align*}
S^{2,3}_{BB} = &E(\hat{a}_0,\hat{b}_0) + E(\hat{a}_0,\hat{b}_1) - E(\hat{a}_0,\hat{b}_2) - E(\hat{a}_0,\hat{b}_3) + \\
			&E(\hat{a}_1,\hat{b}_0) - E(\hat{a}_1,\hat{b}_1) + E(\hat{a}_1,\hat{b}_2) - E(\hat{a}_1,\hat{b}_3) + \\
		  	&E(\hat{a}_2,\hat{b}_0) - E(\hat{a}_2,\hat{b}_1) - E(\hat{a}_2,\hat{b}_2) + E(\hat{a}_2,\hat{b}_3) .
\end{align*}

\noindent Here $\hat{a}_i$ and $\hat{b}_j$ are measurement bases used by analyzers A and B respectively and $E(\hat{a}_i,\hat{b}_j)$ are correlation coefficients. LHV models predict that this inequality is bounded by $S^{2,3}_{BB} \leq 6$, while quantum theory predicts a bound of $S^{2,3}_{BB} \leq 4\sqrt{3} = 6.928$.  A minimal violation of the beautiful Bell inequality requires an entanglement visibility of roughly $87\%$.

\begin{figure}[htp]
\centering
\includegraphics[width=0.75\columnwidth]{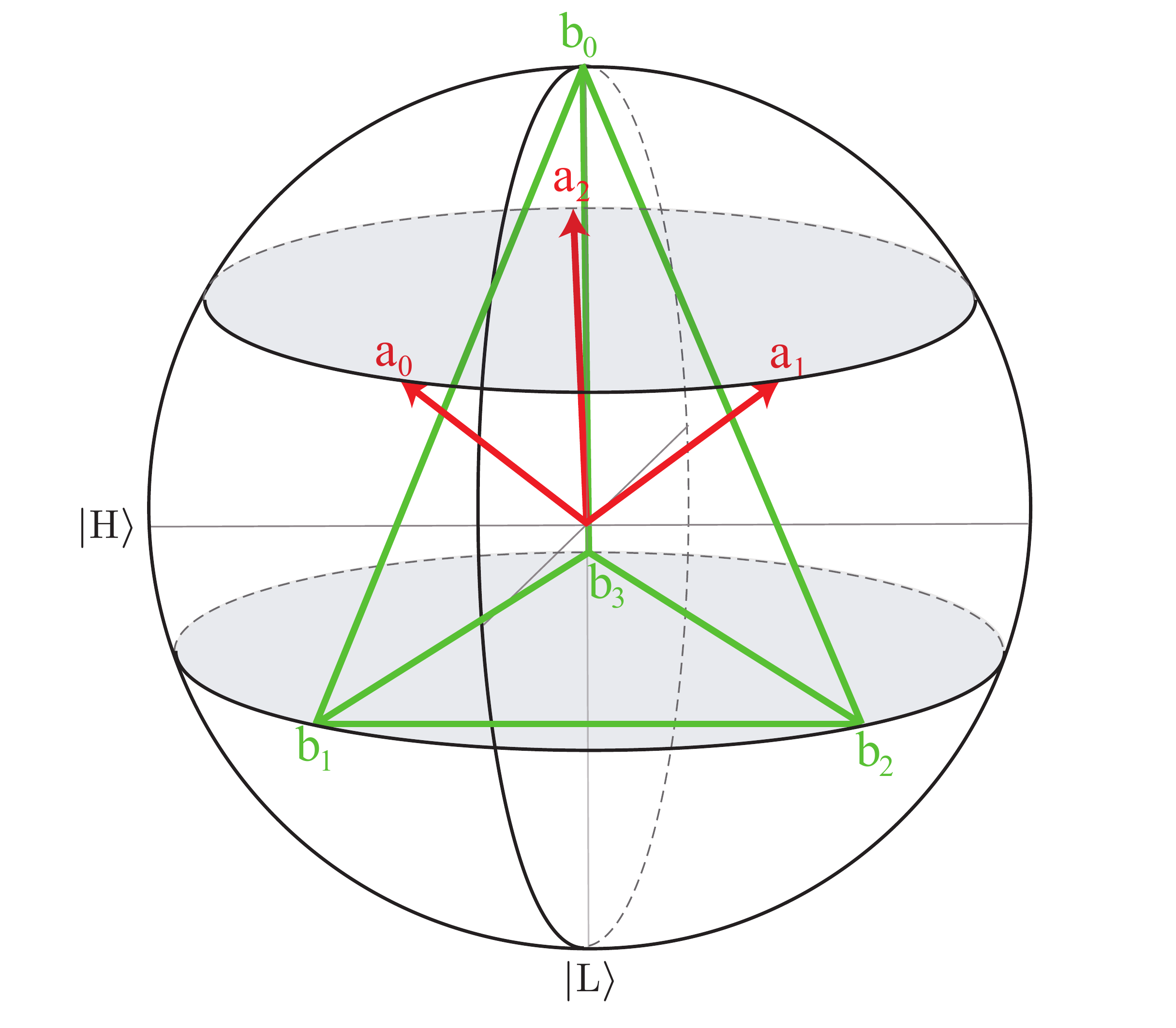}
\caption[Beautiful Bell measurement bases]{\label{fig:belle_bell_settings}{\bf Beautiful Bell measurement bases.} Alice measures in three mutually unbiased bases $\{\hat{a}_0,\hat{a}_1,\hat{a}_2\}$ and Bob measures in bases $\{\hat{b}_0,\hat{b}_1,\hat{b}_2,\hat{b}_3\}$  ~\cite{belle_bell_2}. Only one basis vector (e.g. $a_1$ from $\hat{a}_1=\{a_1,a^\perp_1\}$) from each basis is shown.}
\end{figure}

We measured a value of $S^{2,3}_{BB}=6.67\pm0.08$ (derived from measurement results shown in Table~\ref{tab:belle_bell}), equivalent to a violation of LHV models by over 8 standard deviations.  We are not aware of any previously published experimental violation of the $m=3$, $n=2$ (or higher dimension) ``beautiful'' Bell inequality.

%----------------------------------------------------- Leggett ----------------------------------------------------------------------
\subsection{Leggett inequality}

The Leggett model~\cite{Leggett} differs from deterministic LHV models in that it permits some non-local interactions and makes probabilistic predictions about outcomes of individual measurements.  The Leggett model is interesting because experiments that rule out the LHV models do not automatically rule out NLHV models such as the Leggett model. This model was first experimentally tested in 2007~\cite{GPKBZAZ}.  We tested the 2008 version of the Leggett inequality proposed and first violated by Branciard et al.~\cite{BBGKLLS}, who defined

\begin{equation}
L_3(\varphi) \equiv \frac{1}{3}\sum^3_{i=1}|E(\hat{a}_i,\hat{b}_i)+E(\hat{a}_i,\hat{b}'_i)| .
\end{equation}
\noindent
Here, $E(\hat{a},\hat{b})$ is the correlation function resulting when Alice and Bob measure in pairs of bases separated by angle $\varphi$, as shown in Figure~\ref{fig:leggett_settings}.  The bound provided by the Leggett model for $L_3$ is:

\begin{equation}\label{eq:leggett_bound}
	L_3(\varphi)\leq 2-\frac{2}{3}|\sin\frac{\varphi}{2}| 
\end{equation}

\begin{figure}[htp]
\centering
\includegraphics[width=0.80\columnwidth]{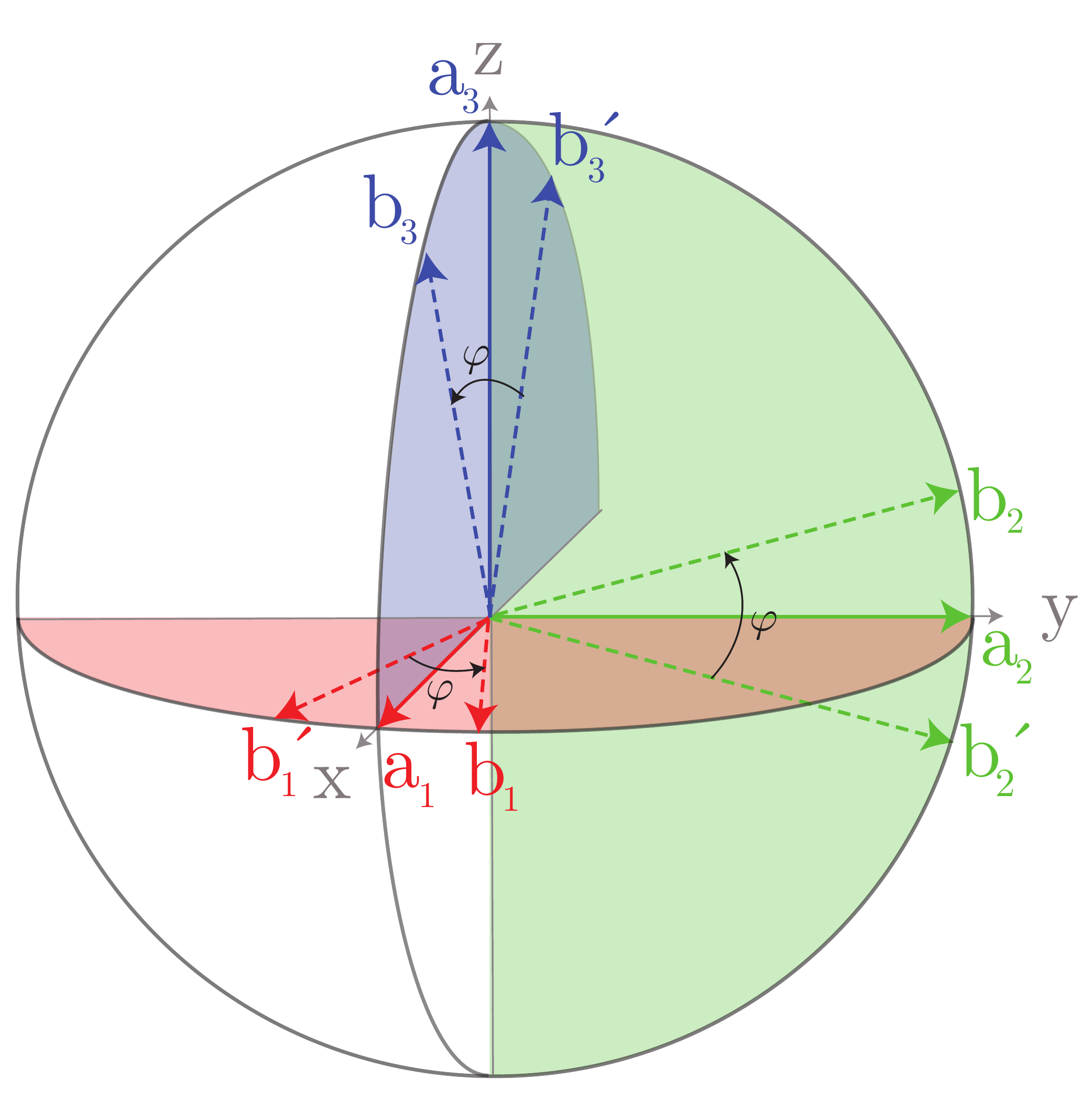}
\caption[Leggett Measurment Settings]{\label{fig:leggett_settings}{\bf Leggett Measurment Settings.} Settings used by Alice (solid lines) and Bob (dashed lines) to test the Leggett inequality.  $b_1$ and $b_1'$ are each separated from $a_1$ by $\frac{\varphi}{2}$, and by $\varphi$ from each other in the XY plane.  Similarly,  $b_2$ and $b_2'$ lie in the YZ plane and  $b_3$ and $b_3'$ are in the XZ plane.}
\end{figure}

\noindent Figure \ref{fig:leggett_results} shows the results we obtained for several different values of $\varphi$.  Each measured point is above the solid red line, which corresponds to the bound of the Leggett model (equation \ref{eq:leggett_bound}) and is therefore a violation of the model.  The maximal violation occurs at $\varphi=40^{\circ}$.  At this setting, the measured value is $L_3=1.82 \pm 0.02$ while the Leggett model is bounded by 1.772 (see Table \ref{tab:Leggett} in the appendix for measurements settings and results for this data point). To our knowledge, this is the first time that the Leggett inequality of the form in~\cite{BBGKLLS} has been violated with photon pairs at non-degenerate wavelengths.  Our result confirms that the specific class of NHLV models described by Leggett is not compatible with experimental observations.

\begin{figure}[htp]
\centering
\includegraphics[width=0.95\columnwidth]{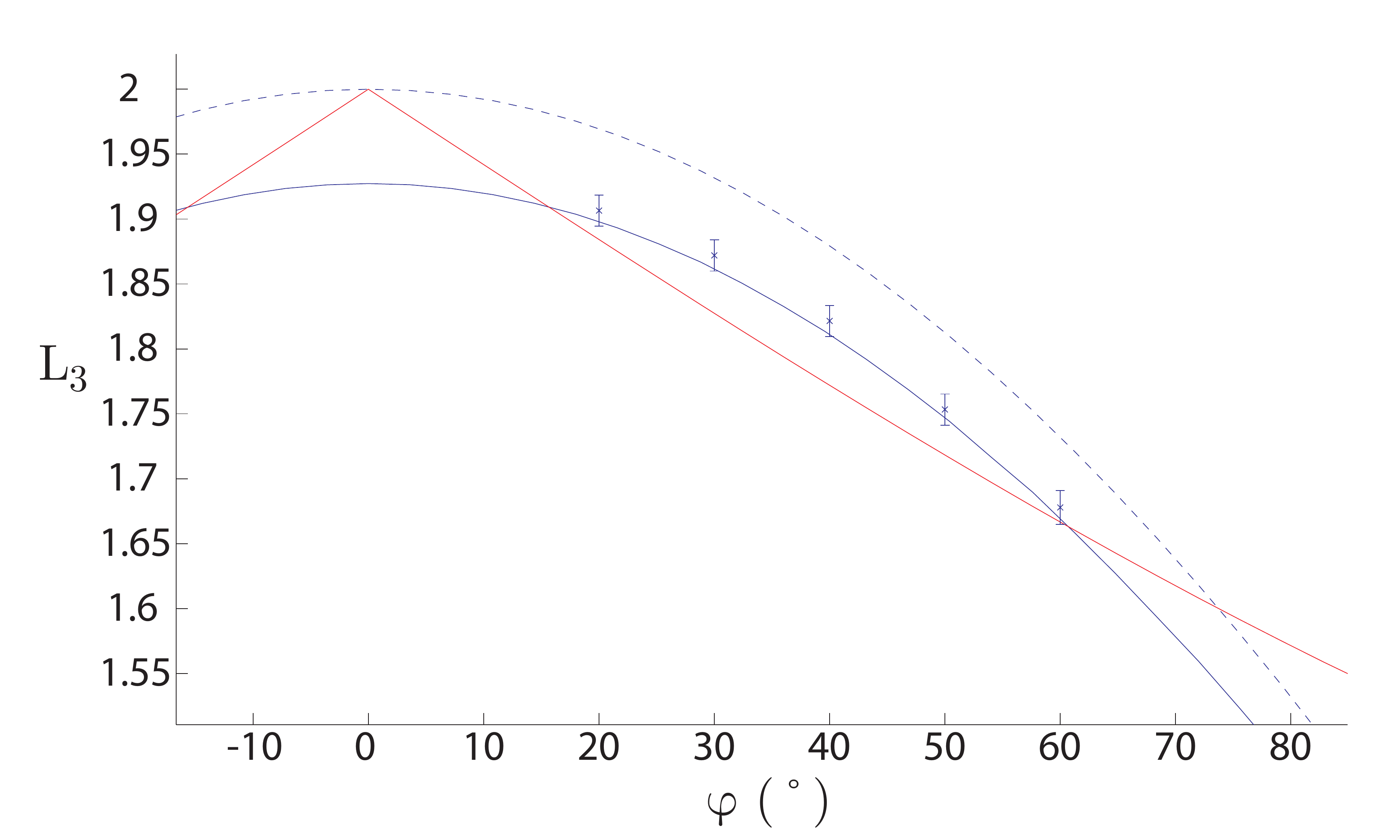}
\caption[Leggett inequality measurement results]{\label{fig:leggett_results}{\bf Leggett inequality measurement results.} Experimentally measured values for $L_3(\varphi)$ are shown versus $\varphi$.  Points with uncertainty bars are experimentally measured values for $L_3(\varphi)$.  The solid red line is the upper bound for the Leggett Model.  Each experimental data point above this line is a violation of the Leggett inequality.  The blue solid line shows predicted $L_3$ values based on a density matrix measured via QST (tangle $\mathcal{T}=0.905$).  The dashed line is the expected $L_3$ value for a perfect $\ket{\Phi^+}$ state.}
\end{figure}

%-------------------------------------------------- Extendibility / Delta -----------------------------------------------------------
%
%One way to avoid testing specific alternatives to quantum theory, such as LHV models, is to bound properties that models must have to be consistent with what we observe in nature.  A recent test was conducted with the source described in this paper that placed a bound on the maximum probability by which any alternative theory could improve on the predictions of quantum theory: $\delta$~\cite{extendibility}. Our results confirmed that LHV and Leggett models have too much predictive power to be consistent with experimental observations.  
%

%--------------------------------------------------- CONCLUSION ---------------------------------------------------------------------

\section{Conclusion}
We have demonstrated a compact and highly flexible source of entangled photon pairs at widely different wavelengths that features high visibility and adjustable tangle. Our source has proved useful for several fundamental tests of quantum theory, namely violations of Bell and Leggett inequalities. It is interesting to note that these tests, which require testing specific inequalities, are not the only way to refute local or certain non-local theories that attempt to explain the origin of quantum correlations. Using the same source, we  recently arrived at the same conclusion based on a more general approach~\cite{extendibility}. More precisely, we ruled out all alternative theories to quantum mechanics, within a causal structure compatible with relativity theory, that improve on quantum mechanical predictions about the outcomes of measurements on maximally entangled particles by more than 16.5\%. In particular, this rules out local and nonlocal hidden variable theories \`a la Bell and Leggett, respectively.

%The ability to adjust wavelength, phase between the $\ket{HH}$ and $\ket{VV}$ components of the state, and the tangle of the state produced gives this design considerable utility for future experiments.   
 
\section*{Acknowledgements}
The authors thank P. Gimby, P. Irwin, and V. Kiselyov for lending material and technical support, and acknowledge funding by NSERC, Alberta Innovates Technology Futures, CFI, AAET and the Killam Trusts.

\newpage
\onecolumngrid
%------------------------------------------------------ SUPPLEMENTS -----------------------------------------------------------------
\newpage
\appendix*
\section{Appendix}

\setcounter{figure}{0} \renewcommand{\thefigure}{A.\arabic{figure}} 
\setcounter{table}{0} \renewcommand{\thetable}{A.\arabic{table}}

\begin{table*}[htp]
\centering
\subtable{
\scriptsize
\centering
\begin{tabular}{c} 
  \\
  \\
  \\
  \\
  \\
  \multirow{-4}{*}{$163.70~^\circ C$}\\ 
\end{tabular}
}
\subtable{
\scriptsize
\centering
\begin{tabular}{r | c  c  c  c  } 
\multicolumn{5}{c}{$\rho_{Re}$} \\
 \multicolumn{1}{r}{} &  \multicolumn{1}{c}{$\bra{HH}$} & \multicolumn{1}{c}{$\bra{HV}$} & \multicolumn{1}{c}{$\bra{VH}$} & \multicolumn{1}{c}{$\bra{VV}$} \\ \cline{2-5}
 $\ket{HH}$ & 0.5062 & 0.0065 & -0.0211 & 0.0742  \\
 $\ket{HV}$ & 0.0065 & 0.0043 & 0.0001 & 0.0196  \\
 $\ket{VH}$ & -0.0211 & 0.0001 & 0.0046 & -0.0102  \\
 $\ket{VV}$ & 0.0742 & 0.0196 & -0.0102 & 0.4849  \\
\end{tabular}
}
\subtable{
\scriptsize
\centering
\begin{tabular}{r | c  c  c  c  } 
\multicolumn{5}{c}{$\rho_{Im}$} \\
 \multicolumn{1}{r}{} &  \multicolumn{1}{c}{$\bra{HH}$} & \multicolumn{1}{c}{$\bra{HV}$} & \multicolumn{1}{c}{$\bra{VH}$} & \multicolumn{1}{c}{$\bra{VV}$} \\ \cline{2-5}
 $\ket{HH}$ & 0.0000 & -0.0110 & -0.0069 & 0.0046  \\
 $\ket{HV}$ & 0.0110 & 0.0000 & 0.0002 & 0.0093  \\
 $\ket{VH}$ & 0.0069 & -0.0002 & 0.0000 & 0.0115  \\
 $\ket{VV}$ & -0.0046 & -0.0093 & -0.0115 & 0.0000  \\
\end{tabular}
}

\subtable{
\scriptsize
\centering
\begin{tabular}{c} 
  \\
  \\
  \\
  \\
  \\
  \multirow{-4}{*}{$164.20~^\circ C$}\\ 
\end{tabular}
}
\subtable{
\scriptsize
\centering
\begin{tabular}{r | c  c  c  c  } 
\multicolumn{5}{c}{$\rho_{Re}$} \\
 \multicolumn{1}{r}{} &  \multicolumn{1}{c}{$\bra{HH}$} & \multicolumn{1}{c}{$\bra{HV}$} & \multicolumn{1}{c}{$\bra{VH}$} & \multicolumn{1}{c}{$\bra{VV}$} \\ \cline{2-5}
 $\ket{HH}$ & 0.4995 & 0.0061 & -0.0217 & 0.1798  \\
 $\ket{HV}$ & 0.0061 & 0.0043 & 0.0008 & 0.0164  \\
 $\ket{VH}$ & -0.0217 & 0.0008 & 0.0059 & -0.0082  \\
 $\ket{VV}$ & 0.1798 & 0.0164 & -0.0082 & 0.4903  \\
\end{tabular}
}
\subtable{
\scriptsize
\centering
\begin{tabular}{r | c  c  c  c  } 
\multicolumn{5}{c}{$\rho_{Im}$} \\
 \multicolumn{1}{r}{} &  \multicolumn{1}{c}{$\bra{HH}$} & \multicolumn{1}{c}{$\bra{HV}$} & \multicolumn{1}{c}{$\bra{VH}$} & \multicolumn{1}{c}{$\bra{VV}$} \\ \cline{2-5}
 $\ket{HH}$ & 0.0000 & -0.0048 & -0.0085 & -0.0091  \\
 $\ket{HV}$ & 0.0048 & 0.0000 & -0.0039 & 0.0125  \\
 $\ket{VH}$ & 0.0085 & 0.0039 & 0.0000 & 0.0092  \\
 $\ket{VV}$ & 0.0091 & -0.0125 & -0.0092 & 0.0000  \\
\end{tabular}
}

\subtable{
\scriptsize
\centering
\begin{tabular}{c} 
  \\
  \\
  \\
  \\
  \\
  \multirow{-4}{*}{$164.70~^\circ C$}\\ 
\end{tabular}
}
\subtable{
\scriptsize
\centering
\begin{tabular}{r | c  c  c  c  } 
\multicolumn{5}{c}{$\rho_{Re}$} \\
 \multicolumn{1}{r}{} &  \multicolumn{1}{c}{$\bra{HH}$} & \multicolumn{1}{c}{$\bra{HV}$} & \multicolumn{1}{c}{$\bra{VH}$} & \multicolumn{1}{c}{$\bra{VV}$} \\ \cline{2-5}
 $\ket{HH}$ & 0.5073 & 0.0027 & -0.0291 & 0.3012  \\
 $\ket{HV}$ & 0.0027 & 0.0049 & 0.0002 & 0.0196  \\
 $\ket{VH}$ & -0.0291 & 0.0002 & 0.0048 & -0.0109  \\
 $\ket{VV}$ & 0.3012 & 0.0196 & -0.0109 & 0.4830  \\
\end{tabular}
}
\subtable{
\scriptsize
\centering
\begin{tabular}{r | c  c  c  c  } 
\multicolumn{5}{c}{$\rho_{Im}$} \\
 \multicolumn{1}{r}{} &  \multicolumn{1}{c}{$\bra{HH}$} & \multicolumn{1}{c}{$\bra{HV}$} & \multicolumn{1}{c}{$\bra{VH}$} & \multicolumn{1}{c}{$\bra{VV}$} \\ \cline{2-5}
 $\ket{HH}$ & 0.0000 & -0.0064 & -0.0043 & -0.0017  \\
 $\ket{HV}$ & 0.0064 & 0.0000 & -0.0038 & 0.0042  \\
 $\ket{VH}$ & 0.0043 & 0.0038 & 0.0000 & 0.0078  \\
 $\ket{VV}$ & 0.0017 & -0.0042 & -0.0078 & 0.0000  \\
\end{tabular}
}

\subtable{
\scriptsize
\centering
\begin{tabular}{c} 
  \\
  \\
  \\
  \\
  \\
  \multirow{-4}{*}{$165.20~^\circ C$}\\ 
\end{tabular}
}
\subtable{
\scriptsize
\centering
\begin{tabular}{r | c  c  c  c  } 
\multicolumn{5}{c}{$\rho_{Re}$} \\
 \multicolumn{1}{r}{} &  \multicolumn{1}{c}{$\bra{HH}$} & \multicolumn{1}{c}{$\bra{HV}$} & \multicolumn{1}{c}{$\bra{VH}$} & \multicolumn{1}{c}{$\bra{VV}$} \\ \cline{2-5}
 $\ket{HH}$ & 0.5249 & 0.0003 & -0.0360 & 0.4007  \\
 $\ket{HV}$ & 0.0003 & 0.0045 & 0.0010 & 0.0210  \\
 $\ket{VH}$ & -0.0360 & 0.0010 & 0.0050 & -0.0111  \\
 $\ket{VV}$ & 0.4007 & 0.0210 & -0.0111 & 0.4656  \\
\end{tabular}
}
\subtable{
\scriptsize
\centering
\begin{tabular}{r | c  c  c  c  } 
\multicolumn{5}{c}{$\rho_{Im}$} \\
 \multicolumn{1}{r}{} &  \multicolumn{1}{c}{$\bra{HH}$} & \multicolumn{1}{c}{$\bra{HV}$} & \multicolumn{1}{c}{$\bra{VH}$} & \multicolumn{1}{c}{$\bra{VV}$} \\ \cline{2-5}
 $\ket{HH}$ & 0.0000 & -0.0033 & -0.0050 & -0.0154  \\
 $\ket{HV}$ & 0.0033 & 0.0000 & -0.0004 & 0.0057  \\
 $\ket{VH}$ & 0.0050 & 0.0004 & 0.0000 & 0.0063  \\
 $\ket{VV}$ & 0.0154 & -0.0057 & -0.0063 & 0.0000  \\
\end{tabular}
}

\subtable{
\scriptsize
\centering
\begin{tabular}{c} 
  \\
  \\
  \\
  \\
  \\
  \multirow{-4}{*}{$165.70~^\circ C$}\\ 
\end{tabular}
}
\subtable{
\scriptsize
\centering
\begin{tabular}{r | c  c  c  c  } 
\multicolumn{5}{c}{$\rho_{Re}$} \\
 \multicolumn{1}{r}{} &  \multicolumn{1}{c}{$\bra{HH}$} & \multicolumn{1}{c}{$\bra{HV}$} & \multicolumn{1}{c}{$\bra{VH}$} & \multicolumn{1}{c}{$\bra{VV}$} \\ \cline{2-5}
 $\ket{HH}$ & 0.5057 & 0.0039 & -0.0301 & 0.4632  \\
 $\ket{HV}$ & 0.0039 & 0.0045 & 0.0025 & 0.0179  \\
 $\ket{VH}$ & -0.0301 & 0.0025 & 0.0052 & -0.0144  \\
 $\ket{VV}$ & 0.4632 & 0.0179 & -0.0144 & 0.4846  \\
\end{tabular}
}
\subtable{
\scriptsize
\centering
\begin{tabular}{r | c  c  c  c  } 
\multicolumn{5}{c}{$\rho_{Im}$} \\
 \multicolumn{1}{r}{} &  \multicolumn{1}{c}{$\bra{HH}$} & \multicolumn{1}{c}{$\bra{HV}$} & \multicolumn{1}{c}{$\bra{VH}$} & \multicolumn{1}{c}{$\bra{VV}$} \\ \cline{2-5}
 $\ket{HH}$ & 0.0000 & -0.0028 & -0.0047 & -0.0264  \\
 $\ket{HV}$ & 0.0028 & 0.0000 & -0.0005 & 0.0045  \\
 $\ket{VH}$ & 0.0047 & 0.0005 & 0.0000 & 0.0068  \\
 $\ket{VV}$ & 0.0264 & -0.0045 & -0.0068 & 0.0000  \\
\end{tabular}
}

\subtable{
\scriptsize
\centering
\begin{tabular}{c} 
  \\
  \\
  \\
  \\
  \\
  \multirow{-4}{*}{$166.20~^\circ C$}\\ 
\end{tabular}
}
\subtable{
\scriptsize
\centering
\begin{tabular}{r | c  c  c  c  } 
\multicolumn{5}{c}{$\rho_{Re}$} \\
 \multicolumn{1}{r}{} &  \multicolumn{1}{c}{$\bra{HH}$} & \multicolumn{1}{c}{$\bra{HV}$} & \multicolumn{1}{c}{$\bra{VH}$} & \multicolumn{1}{c}{$\bra{VV}$} \\ \cline{2-5}
 $\ket{HH}$ & 0.5125 & 0.0111 & -0.0305 & 0.4770  \\
 $\ket{HV}$ & 0.0111 & 0.0048 & 0.0015 & 0.0197  \\
 $\ket{VH}$ & -0.0305 & 0.0015 & 0.0051 & -0.0191  \\
 $\ket{VV}$ & 0.4770 & 0.0197 & -0.0191 & 0.4775  \\
\end{tabular}
}
\subtable{
\scriptsize
\centering
\begin{tabular}{r | c  c  c  c  } 
\multicolumn{5}{c}{$\rho_{Im}$} \\
 \multicolumn{1}{r}{} &  \multicolumn{1}{c}{$\bra{HH}$} & \multicolumn{1}{c}{$\bra{HV}$} & \multicolumn{1}{c}{$\bra{VH}$} & \multicolumn{1}{c}{$\bra{VV}$} \\ \cline{2-5}
 $\ket{HH}$ & 0.0000 & 0.0000 & -0.0029 & -0.0281  \\
 $\ket{HV}$ & 0.0000 & 0.0000 & 0.0010 & 0.0028  \\
 $\ket{VH}$ & 0.0029 & -0.0010 & 0.0000 & 0.0015  \\
 $\ket{VV}$ & 0.0281 & -0.0028 & -0.0015 & 0.0000  \\
\end{tabular}
}

\subtable{
\scriptsize
\centering
\begin{tabular}{c} 
  \\
  \\
  \\
  \\
  \\
  \multirow{-4}{*}{$166.70~^\circ C$}\\ 
\end{tabular}
}
\subtable{
\scriptsize
\centering
\begin{tabular}{r | c  c  c  c  } 
\multicolumn{5}{c}{$\rho_{Re}$} \\
 \multicolumn{1}{r}{} &  \multicolumn{1}{c}{$\bra{HH}$} & \multicolumn{1}{c}{$\bra{HV}$} & \multicolumn{1}{c}{$\bra{VH}$} & \multicolumn{1}{c}{$\bra{VV}$} \\ \cline{2-5}
 $\ket{HH}$ & 0.5076 & 0.0084 & -0.0317 & 0.4396  \\
 $\ket{HV}$ & 0.0084 & 0.0052 & 0.0007 & 0.0226  \\
 $\ket{VH}$ & -0.0317 & 0.0007 & 0.0045 & -0.0157  \\
 $\ket{VV}$ & 0.4396 & 0.0226 & -0.0157 & 0.4827  \\
\end{tabular}
}
\subtable{
\scriptsize
\centering
\begin{tabular}{r | c  c  c  c  } 
\multicolumn{5}{c}{$\rho_{Im}$} \\
 \multicolumn{1}{r}{} &  \multicolumn{1}{c}{$\bra{HH}$} & \multicolumn{1}{c}{$\bra{HV}$} & \multicolumn{1}{c}{$\bra{VH}$} & \multicolumn{1}{c}{$\bra{VV}$} \\ \cline{2-5}
 $\ket{HH}$ & 0.0000 & -0.0032 & -0.0052 & -0.0227  \\
 $\ket{HV}$ & 0.0032 & 0.0000 & -0.0019 & 0.0040  \\
 $\ket{VH}$ & 0.0052 & 0.0019 & 0.0000 & 0.0069  \\
 $\ket{VV}$ & 0.0227 & -0.0040 & -0.0069 & 0.0000  \\
\end{tabular}
}

\caption[Tangle versus Spectral Overlap Density Matrices]{\label{tab:tangle_vs_temp}{\bf Tangle versus Spectral Overlap Density Matrices.} Density matrices measured as $\ket{HH}$ SPDC crystal temperature (shown above) was varied. Phase was adjusted for maximal fidelity to a $\ket{\Phi^+}$ (positive values for off-diagonal terms $\ket{HH}\bra{VV}$ and $\ket{VV}\bra{HH}$) or $\ket{\Phi^-}$ (negative values for off-diagonal terms) state.  The $\ket{VV}$ SPDC crystal temperature was kept at a constant $165.70~^\circ C$.}
\end{table*}
\normalsize

\newpage

\begin{table*}[htp]
  \begin{center}
  \begin{tabular}{ c | r  r | c |  r  r  } 
  \toprule
  \multicolumn{1}{c}{Bases} & \multicolumn{1}{|c}{$E(\hat{a}_i,\hat{b}_j)$} & \multicolumn{1}{c}{$\Delta E(\hat{a}_i,\hat{b}_j)$} & \multicolumn{1}{|c}{Bases} & \multicolumn{1}{c}{$E(\hat{a}_i,\hat{b}_j)$} & \multicolumn{1}{c}{$\Delta E(\hat{a}_i,\hat{b}_j)$}\\ \hline
$\{\hat{a}_0,\hat{b}_0\}$ & 0.5742  & 0.0061 & $\{\hat{a}_1,\hat{b}_2\}$ & 0.5763  & 0.0060  \\ 
$\{\hat{a}_0,\hat{b}_1\}$ & 0.5247  & 0.0062 & $\{\hat{a}_1,\hat{b}_3\}$ & -0.5833 & 0.0061  \\ 
$\{\hat{a}_0,\hat{b}_2\}$ & -0.5641 & 0.0062 & $\{\hat{a}_2,\hat{b}_0\}$ & 0.6124  & 0.0061  \\ 
$\{\hat{a}_0,\hat{b}_3\}$ & -0.5678 & 0.0061 & $\{\hat{a}_2,\hat{b}_1\}$ & -0.6255 & 0.0061  \\ 
$\{\hat{a}_1,\hat{b}_0\}$ & 0.5446  & 0.0061 & $\{\hat{a}_2,\hat{b}_2\}$ & -0.5039 & 0.0061  \\ 
$\{\hat{a}_1,\hat{b}_1\}$ & -0.5307 & 0.0061 & $\{\hat{a}_2,\hat{b}_3\}$ & 0.4645  & 0.0061  \\ 
\hline
  \bottomrule
  \end{tabular}
  \normalsize
  \caption[Beautiful Bell Measurement Settings and Data]{\label{tab:belle_bell}{\bf Beautiful Bell Measurement Settings and Data.} This table shows raw data collected to find $S_{BB}=6.67\pm0.08 > 6$. $E(\hat{a}_i,\hat{b}_j)$ is the correlation coefficient measured using bases $\hat{a}_i$ and $\hat{b}_j$.  Four coincidence measurements (not shown) consisting of 40 second samples were recorded for each correlation coefficient.  Uncertainties are derived from Poissonian statistics.}
  \end{center}
\end{table*}

\begin{table*}[htp]
  \begin{center}
  \begin{tabular}{ c | r  r  } 
  \toprule
  \multicolumn{1}{c}{Bases} & \multicolumn{1}{|c}{$E(\hat{a}_i,\hat{b}_j)$} & \multicolumn{1}{c}{$\Delta E(\hat{a}_i,\hat{b}_j)$} \\ \hline
$\{\hat{a}_1,\hat{b}_1\}$ & 0.9083  & 0.0057 \\ 
$\{\hat{a}_1,\hat{b}_1'\}$ & 0.8919  & 0.0057 \\ 
$\{\hat{a}_2,\hat{b}_2\}$ & -0.9081 & 0.0038 \\ 
$\{\hat{a}_2,\hat{b}_2'\}$ & -0.8972 & 0.0059 \\ 
$\{\hat{a}_3,\hat{b}_3\}$ & 0.9199  & 0.0059 \\ 
$\{\hat{a}_3,\hat{b}_3'\}$ & 0.9391 & 0.0060 \\ 
\hline
  \bottomrule
  \end{tabular}
  \normalsize
  \caption[Leggett Inequality Data]{\label{tab:Leggett}{\bf Leggett Inequality Data ($\varphi=40^\circ$).} This table shows correlation coefficients, $E(\hat{a}_i,\hat{b}_j)$, measured between bases $\hat{a}_i$ and $\hat{b}_j$ respectively to find $L_3=1.82 \pm 0.02>1.772$ for $\varphi=40^\circ$. Data collection time for each point was 40 seconds.  Uncertainties are derived from Poissonian statistics.}
  \end{center}
\end{table*}

%------------------------------------------------------ BIBLIOGRAPHY ----------------------------------------------------------------
\clearpage
\twocolumngrid

\end{document}